# Valley polarization and valleyresistance in monolayer transition metal dichalcogenides superlattice

Hui-Ying Mu[1], Yi-Tong Yao[1], Jie-Ru Li[1], Guo-Cai Liu[1], Chao He[1], Ying-Jie Sun[1], Guang Yang[1], Xing-Tao An[1*], Yong-Zhe Zhang[2*], Jian-Jun Liu[3*]

[1]School of Science, Hebei University of Science and Technology, Shijiazhuang, Hebei 050018, China
[2]College of Materials Science and Engineering, Beijing University of Technology, No. 100 Pingleyuan Chaoyang District, Beijing 100124, China.
[3]Physics Department, Shijiazhuang University, Shijiazhuang, Hebei 050035, China

*Correspondence to: anxt2005@163.com, yzzhang@bjut.edu.cn, liujj@hebtu.edu.cn

**Manipulating the valley degree of freedom to encode information for potential valleytronic devices has ignited a new direction in solid-state physics. A significant, fundamental challenge in the field of valleytronics is how to generate and regulate valley-polarized currents by practical ways. Here, we discover a new mechanism of producing valley polarization in a monolayer transition metal dichalcogenides superlattice, in which valley-resolved gaps are formed at the supercell Brillouin zone boundaries and centers due to the intervalley scattering. When the energy of the incident electron is in the gaps, the available states are valley polarized, thus providing a valley-polarized current from the superlattice. We show that the direction and strength of the valley polarization may further be tuned by varying the potential applied the superlattice. The transmission can have a net valley polarization of 55% for a 4-period heterojunction. Moreover, such two valley filters in series may function as an electrostatically controlled giant valleyresistance device, representing a zero magnetic field counterpart to the familiar giant magnetoresistance device.**

In the electronic band structure of a crystalline solid, a local minima in the conduction band or the valence band is referred to as a valley. Two-dimensional hexagonal materials, such as graphene and monolayer transition metal dichalcogenides have two inequivalent degenerate valleys that are well separated in momentum space at the first Brillouin zone. The two valleys can exert the same effect as the electron spin in storing and carrying information, which leads to an emerging device concept known as valleytronics. Creating a valley polarization, compelling electrons to selectively populate one valley or another, is the prerequisite for valleytronic applications.

There were tremendous works devoted to generate valley-polarized currents by using edges[1], doping[2], line defects[3], lattice strains[4-6], or valley-dependent trigonal warping of the dispersions[7]. Meanwhile, the spin-like properties of the valley, such as the valley Hall effect[8-14], the valley magnetic response[15-18], and the valley optical selection rules[9,19], allow us to manipulate the valley by electromagnetic[20-26] or optical field[27-30] as we control the spin. These valley control schemes rely on specific materials properties or the inversion symmetry breaking in the two-dimensional hexagonal lattices. In fact, the valley-polarized current can be pumped simply by the ubiquitous nonmagnetic disorders, not relying on any specific material property, because the intervalley backscattering with the distinct momentum transfers cause a net transfer of population from one valley to another[31]. Based on the intra- and inter-valley scatterings, the switchable valley filter and valley source functions may be realized from the lateral junctions in graphene and transition metal dichalcogenides[32]. However, the exploration towards practical valley polarization for potential valleytronic devices remains an ongoing challenge.

Here, we discover a new mechanism to generate electrically tunable valley polarization based on the intervalley scattering in a monolayer transition metal dichalcogenides lateral superlattice. The intravalley backscattering of the charge carriers tunneling through such a semiconductor superlattice can form the conventional minibands and minigaps. On the other hand, due to the distinct momentum transfers on a finite Fermi surface, the intervalley backscattering can cause the novel valley-resolved minigaps. In general, the intervalley resonant tunneling phenomenon is hard to observe and is neglected due to the large distance between the two valleys. For the monolayer transition metal dichalcogenides lateral superlattice, however, two kinds of different multiple intervalley backscattering result in valley-resolved gaps at the supercell Brillouin zone boundaries and centers, respectively. Therefore, the superlattice can be used as a valley filter. The conductance can have a net valley polarization of 55% for a 4-period superlattice. This valley polarization arising from its momentum nature, rather than the spin-like properties, can be tuned by adjusting the incident energy or the external potential. Remarkably two superlattices in series may function as an electrostatically controlled valleyresistance device.

To understand the electronic properties of the monolayer transition metal dichalcogenides superlattices, let us analyze the situation of the electron tunneling through a single barrier of width $W$ on monolayer transition metal dichalcogenides first. In these two-valley materials with the parabolic dispersion relation, there are two kinds of intervalley resonant tunneling conditions. One is $(K-q)W = n\pi$ and the other is $(K+q)W = n\pi$, as shown in Fig. 1 (a), where $q$ is measured from the valley center $\pm K$ and $n$ is integer. Therefore, the intervalley resonant tunneling energies between the states $|K-q\rangle$ and $|-(K-q)\rangle$ can be written as

$$E_+ = (K + n\pi/W)^2 \hbar^2 / 2m, \tag{1}$$

where $m$ is the electron effective mass. And the intervalley resonant tunneling energies between the states $|K+q\rangle$ and $|-(K+q)\rangle$ can be written as

$$E_- = (K - n\pi/W)^2 \hbar^2 / 2m. \tag{2}$$

On the other hand, the intravalley resonant tunneling conditions of the two valleys are identical, i.e., $qW = n\pi$ and the intravalley resonant tunneling energies are

$$E_0 = (n\pi/W)^2 \hbar^2 / 2m. \tag{3}$$

As the number of the barriers or the period of the heterojunction increases, two kinds of intervalley resonant levels will evolve into two sets of valley-dependent minibands and the corresponding minigaps appear alternately at different energy regions.

In order to prove the tunable electrically valley polarization in the monolayer transition metal dichalcogenides lateral superlattice, the valley-dependent electron transport along zigzag direction in a monolayer $MoS_2/WS_2$ superlattice is studied as example. A schematic depiction of the device based on $MoS_2/WS_2$ is shown in Fig. 1(b). We adopt the three-band tight-binding model, considering hoppings between nearest-neighbour $Mo-d_{z^2}$, $d_{xy}$, and $d_{x^2-y^2}$ orbitals[33]. The monolayer transition metal dichalcogenides can be described by the following Hamiltonian:

$$H = \sum_i \varepsilon_i c_i^\dagger c_i + \sum_{<i,i'>} \mathbf{t}(\vec{r}_{i'}-\vec{r}_i) c_i^\dagger c_{i'} + h.c., \qquad (4)$$

where $c_i^\dagger$ is the creation operator for electron on site $i$. $<i,i'>$ run over all the nearest neighbor hopping sites. $\vec{r}_{i'}-\vec{r}_i$ is the six nearest-neighbour vectors $\vec{R}_1$ to $\vec{R}_6$. The on-site energy $\varepsilon_i$ and the nearest-neighbor hopping $\mathbf{t}(\vec{r}_{i'}-\vec{r}_i)$ are $3\times 3$ matrices and given by

$$\varepsilon_i = \begin{pmatrix} \varepsilon_1 & 0 & 0 \\ 0 & \varepsilon_2 & 0 \\ 0 & 0 & \varepsilon_2 \end{pmatrix},$$

$$\mathbf{t}(\vec{R}_1) = \begin{pmatrix} t_0 & t_1 & t_2 \\ -t_1 & t_{11} & t_{12} \\ t_2 & -t_{12} & t_{22} \end{pmatrix},$$

$$\mathbf{t}(\vec{R}_2) = \begin{pmatrix} t_0 & \frac{1}{2}(t_1-\sqrt{3}t_2) & -\frac{1}{2}(\sqrt{3}t_1+t_2) \\ -\frac{1}{2}(t_1+\sqrt{3}t_2) & \frac{1}{4}(t_{11}+3t_{22}) & -\frac{\sqrt{3}}{4}(t_{11}-t_{22})-t_{12} \\ \frac{1}{2}(\sqrt{3}t_1-t_2) & -\frac{\sqrt{3}}{4}(t_{11}-t_{22})+t_{12} & \frac{1}{4}(3t_{11}+t_{22}) \end{pmatrix},$$

$$\mathbf{t}(\bar{R}_3) = \begin{pmatrix} t_0 & -\frac{1}{2}(t_1 - \sqrt{3}t_2) & -\frac{1}{2}(\sqrt{3}t_1 + t_2) \\ \frac{1}{2}(t_1 + \sqrt{3}t_2) & \frac{1}{4}(t_{11} + 3t_{22}) & \frac{\sqrt{3}}{4}(t_{11} - t_{22}) + t_{12} \\ \frac{1}{2}(\sqrt{3}t_1 - t_2) & \frac{\sqrt{3}}{4}(t_{11} - t_{22}) - t_{12} & \frac{1}{4}(3t_{11} + t_{22}) \end{pmatrix},$$

$\mathbf{t}(\bar{R}_4) = \mathbf{t}(\bar{R}_1)^\dagger$, $\mathbf{t}(\bar{R}_5) = \mathbf{t}(\bar{R}_2)^\dagger$, $\mathbf{t}(\bar{R}_6) = \mathbf{t}(\bar{R}_3)^\dagger$.

In order to simulate two-dimensional bulk $MoS_2/WS_2$ superlattice we impose periodic boundary conditions on nanoribbons of width $D$ oriented with zigzag carbon chains along the $x$ direction. This is done via modifying the hopping between atomic sites connected through the periodic boundary conditions by a Bloch phase factor $e^{ik_y D}$ with the component of the electron's momentum $k_y$ perpendicular to the nanoribbon, as described in Ref. [34]. Consider a $MoS_2/WS_2$ superlattice connected to the outer world by two semi-infinite pristine $MoS_2$ leads, the valley-resolved transport properties of $MoS_2/WS_2$ superlattice are calculated with the tight-binding Hamiltonian in Eq. (4), using a recursive Green's function technique[35].

Let us focus on the bottom of the conduction band, where the spin-orbit coupling of the system is weak and can be ignored. The conduction band offset of $MoS_2$ and $WS_2$ is about 150meV, which can be obtained by calculating the band structure of monolayer $MoS_2$ and $WS_2$ described Eq. (4). The superlattice potential is shown in the lower panel in Fig. 1 (b). The conduction band bottom of monolayer $MoS_2$ is defined as the zero of energy. Fig. 1 (c) shows an example of the superlattice energy bands, with $L = 10a$, $W = 5a$, $a$ being the lattice constant of $MoS_2$. The growth of sub–2-nm quantum-well arrays in semiconductor monolayers has been implemented experimentally[36]. There is a sizable bandgap δ corresponding to the multiple intravalley reflection. In addition, it is interesting that two gaps $δ_+$, and $δ_-$ induced by two kinds of intervalley reflection, respectively, are opened at the boundary ($k_x = \pm\pi/L$) and center ($k_x = 0$) of the supercell Brillouin zone [indicated by the arrows in Fig. 1 (c)]. Fig. 1 (d) is a zoom-in of the band dispersion with a cut at $k_y = 0$, which includes the lowest two subbands. The two valley-resolved gaps are highlighted by the red and blue shadows. Within the gaps $\delta_+$ and $\delta_-$, the states are valley polarized. The intervalley scattering makes possible energy windows for valley-polarized transport in the superlattice.

Fig. 1 (e) plots the two lowest allowed energy bands as functions of the barrier width $W$ for a superlattice, fixing well width $L - W = 5a$. If the quantum wells are separated well beyond the characteristic decay length of a particular state in the barrier regions, for example $L > 25a$, then they are expected to behave as isolated single wells. The valley degeneracy of the states in such a system of uncoupled quantum wells is split by the intervalley coupling due to the potential confinement. The valley splitting can reach to 12meV, as shown in Fig. 1 (e). When the barriers thickness between the wells is decreased, the envelope wavefunctions in neighbouring wells overlap. As expected, the discrete valley splitting states now become a continuum of states (valley-resolved minibands). Each group of the minibands separated by gaps $\delta$ has a pair of valley-resolved minigaps $\delta_+$ and $\delta_-$.

The valley-resolved minigaps as functions of the barrier width are shown in Fig. 1 (f). The magnitudes of the gaps are nonmonotonic with the increase $W$, but rather have oscillations which are out-of-phase between the two valley-resolved gaps. The maxima values of the gaps $\delta_+$ and $\delta_-$ are respectively determined by the maximal intervalley backscattering conditions. As the barrier width increases, the two gaps tend to the same value because the minibands shrink in width back to the discrete states of the individual wells.

Fig. 2 shows the calculated valley-conserved and valley-flip transmission and reflection coefficients under normal incidence ( $k_y = 0$ ), for 2, 3, 50-period superlattices with $L = 10a$, and $W = 5a$. In the case of a superlattice with a small number of periods, transmission exhibits strong resonances. Remarkably, the peak for resonant tunneling structure (*N*=2) splits in a pair of valley-dependent peaks compared to the traditional semiconductor double barriers. This result can be understood qualitatively in a tight-binding scheme as arising from the valley-resolved quasibound states in the well [see Fig. 1 (e)]. As the periods increase, these valley-resolved quasibound states in neighbouring wells overlap and produce split groups of valley-resolved transmission resonances. Each group of valley-resolved resonances evolves into a continuous valley-resolved miniband in the limit of a many-period superlattice. The valley-resolved miniband splitting is more pronounced for higher energy. This fact reflects the larger penetration of the upper states envelope functions into the $WS_2$ due to the smaller effective barrier.

In the energy gap $\delta$ shown in Fig. 2 (b), the reflection of the system is mainly contributed by the intravalley reflection coefficients, $R_{-k,-k}$ and $R_{k,k}$. While in the the valley-resolved gaps $\delta_+$ and $\delta_-$, the reflection of the system consists mainly of intervalley reflection coefficients, $R_{-k,k}$ or $R_{k,-k}$. These results are in perfect agreement with the previous theoretical analysis. In the shaded energy window that corresponds to the valley-resolved gap $\delta_+$, especially at higher energy region shown in Fig. 2, we find remarkable valley-conserving transmission for incidence in valley $-K$, and obvious valley-flip reflection for incidence in valley K. The energy window $\delta_-$ exhibits the same behavior with opposite valley polarization for the allowed/blocked transmission.

The valley-resolved conductance for the electron current through such a structure can be calculated by summing $T_{\tau,\tau'}(k_y)$ with respect to the transverse momentum, $G_{\tau,\tau'} = \frac{2e^2}{h}\sum_{k_y} T_{\tau,\tau'}(k_y)$, where $\tau,\tau' = \pm K$. The range of the transverse momentum for the sum is $[-\frac{\pi}{D},\frac{\pi}{D}]$ with the interval $\Delta k_y = 2\pi/L_y$, where $L_y$ is the transverse width of the superlattice. The valley polarization can be defined as $P_V = (G_{-K,-K} + G_{K,-K} - G_{K,K} - G_{-K,K})/G_{sum}$, where $G_{sum} = G_{K,K} + G_{-K,-K} + G_{K,-K} + G_{-K,K}$ is the overall conductance. Fig. 3 (a) shows the calculated valley-resolved conductances as functions of the incident energy, for a 20-period superlattice with $L_y = 120\sqrt{3}a \approx 65nm$, $L = 10a$, and $W = 5a$. Due to the same momentum transfers, the two valley-flip conductances are always identical i.e. $G_{K,-K} = G_{-K,K}$, even for multimodal transport. For a superlattice of finite width, the two valley-conserved conductances $G_{K,K}$ and $G_{-K,-K}$ are also different greatly in the valley-resolved gaps $\delta_+$ and $\delta_-$. Therefore, this leads to a pronounced valley polarization almost reaching

~ 60%, as shown in Fig. 3 (b). Moreover, it is found that the direction and strength of valley polarization can be controlled by the incident energy.

One of the key factors to produce valley-polarized currents in practice is the period of the superlattice in the scattering region. Fig. 3 (c) shows the valley polarization $P_v$ as a function of the period $N$ of the superlattice with $L = 10a$, and $W = 5a$ for the various incident energies in the valley-resolved gaps $\delta_+$. A few periods are sufficient to achieve high efficiency valley filter as Fig. 3 (c) has shown. For example, it only takes four periods to realize almost 55% valley polarization in the first valley-resolved gap $\delta_+$. More superlattice periods are required in order to realize higher valley polarization in the second valley-resolved gap $\delta_+$.

The valley polarization of the superlattice can be inverted by locally raising the valley-resolved gaps by means of a gate voltage, such that the incident energy lies in the gaps $\delta_+$ and $\delta_-$. Two sets of superlattices in series, shown in Fig. 4 (a), can block the current if they have the opposite valley polarization, demonstrating that two superlattices can operate as a 'giant valleyresistance' (GVR) device—a purely electronic counterpart of the giant magnetoresistance device. The GVR can be defined as

$$\text{GVR} = \frac{G_{sum,P} - G_{sum,AP}}{G_{sum,AP}},$$

where $G_{sum,P}$ and $G_{sum,AP}$ are the total conductances corresponding to the valley parallel and antiparallel configurations, respectively. The operation of the valleyresistance device is demonstrated in Fig. 4 (b). The current is blocked as the chemical potential of one of superlattice $\mu_R = \mu_L + eV_g$ is adjusted into the gap $\delta_-$ with the other one $\mu_L$ fixed in the gap $\delta_+$, so that the device contains two valley filters of opposite polarization in series. Such a device can easily give values of GVR exceeding 200%. We anticipate that the experimental realization of this device have practical significance in applications for future valleytronic device.

**Acknowledgment**

This work is mainly supported by Hebei Funds for Distinguished Young Scientists

(No. A201808076) and Hebei Hundred Excellent Innovative Talents (No. SLRC2017035). Y. S. is also supported by the Natural Science Foundation of Hebei Province (Grant No. B2019208201) and NSFC 21903019.

**References**

[1] A. Rycerz, J. Tworzydlo, and C. W. J. Beenakker, Nat. Phys. **3**, 172 (2007).

[2] N. Singh and U. Schwingenschlögl, Adv. Mater. **29**, 1600970 (2017).

[3] D. Gunlycke and C. T. White, Phys. Rev. Lett. **106**, 136806 (2011).

[4] T. Low and F. Guinea, Nano Lett. **10**, 3551 (2010).

[5] Z. Wu, F. Zhai, F. M. Peeters, H. Q. Xu, and K. Chang, Phys. Rev. Lett. **106**, 176802 (2011).

[6] A. Georgi *et al.*, Nano Lett. **17**, 2240 (2017).

[7] H. Yu, Y. Wu, G.-B. Liu, X. Xu, and W. Yao, Phys. Rev. Lett. **113**, 156603 (2014).

[8] D. Xiao, W. Yao, and Q. Niu, Phys. Rev. Lett. **99**, 236809 (2007).

[9] D. Xiao, G.-B. Liu, W. Feng, X. Xu, and W. Yao, Phys. Rev. Lett. **108**, 196802 (2012).

[10] R. V. Gorbachev *et al.*, Science **346**, 448 (2014).

[11] M. Sui *et al.*, Nat. Phys. **11**, 1027 (2015).

[12] Y. Shimazaki, M. Yamamoto, I. V. Borzenets, K. Watanabe, T. Taniguchi, and S. Tarucha, Nat. Phys. **11**, 1032 (2015).

[13] K. F. Mak, K. L. McGill, J. Park, and P. L. McEuen, Science **344**, 1489 (2014).

[14] K. Komatsu, Y. Morita, E. Watanabe, D. Tsuya, K. Watanabe, T. Taniguchi, and S. Moriyama, Sci. Adv. **4**, eaaq0194 (2018).

[15] A. Srivastava, M. Sidler, A. V. Allain, D. S. Lembke, A. Kis, and A. Imamoglu, Nat. Phys. **11**, 141 (2015).

[16] G. Aivazian *et al.*, Nat. Phys. **11**, 148 (2015).

[17] Y. Li *et al.*, Phys. Rev. Lett. **113**, 266804 (2014).

[18] D. MacNeill, C. Heikes, K. F. Mak, Z. Anderson, A. Kormanyos, V. Zolyomi, J. Park, and D. C. Ralph, Phys. Rev. Lett. **114**, 037401 (2015).

[19] W. Yao, D. Xiao, and Q. Niu, Phys. Rev. B **77**, 235406 (2008).

[20] Y. Ye *et al.*, Nat. Nanotech. **11**, 598 (2016).

[21] D. Zhong *et al.*, Sci. Adv. **3**, e1603113 (2017).

[22] Q. Zhang, S. A. Yang, W. Mi, Y. Cheng, and U. Schwingenschlögl, Adv. Mater. **28**, 959 (2016).

[23] P. Nagler *et al.*, Nat. Commun. **8**, 1551 (2017).

[24] P. Back, M. Sidler, O. Cotlet, A. Srivastava, N. Takemura, M. Kroner, and A. Imamoğlu, Phys. Rev. Lett. **118**, 237404 (2017).

[25] A. Kundu, H. A. Fertig, and B. Seradjeh, Phys. Rev. Lett. **116**, 016802 (2016).

[26] C. Jin *et al.*, Science **360**, 893 (2018).

[27] H. Zeng, J. Dai, W. Yao, D. Xiao, and X. Cui, Nat. Nanotech. **7**, 490 (2012).

[28] K. F. Mak, K. He, J. Shan, and T. F. Heinz, Nat. Nanotech. **7**, 494 (2012).

[29] T. Cao *et al.*, Nat. Commun. **3**, 887, 887 (2012).

[30] A. M. Jones *et al.*, Nat. Nanotech. **8**, 634 (2013).

[31] X.-T. An, J. Xiao, M. W. Y. Tu, H. Yu, V. I. Fal'ko, and W. Yao, Phys. Rev. Lett. **118**, 096602 (2017).

[32] M. W.-Y. Tu and W. Yao, 2D Mater. **4**, 025109 (2017).

[33] G.-B. Liu, W.-Y. Shan, Y. Yao, W. Yao, and D. Xiao, Phys. Rev. B **88**, 085433 (2013).

[34] M.-H. Liu, J. Bundesmann, and K. Richter, Phys. Rev. B **85**, 085406 (2012).

[35] T. Ando, Phys. Rev. B **44**, 8017 (1991).

[36] W. Zhou, Y.-Y. Zhang, J. Chen, D. Li, J. Zhou, Z. Liu, M. F. Chisholm, S. T. Pantelides, and K. P. Loh, Sci. Adv. **4**, eaap9096 (2018).

[23] P. Nagler *et al.*, Nat. Commun. **8**, 1551 (2017).

[24] P. Back, M. Sidler, O. Cotlet, A. Srivastava, N. Takemura, M. Kroner, and A. Imamoğlu, Phys. Rev. Lett. **118**, 237404 (2017).

[25] A. Kundu, H. A. Fertig, and B. Seradjeh, Phys. Rev. Lett. **116**, 016802 (2016).

[26] C. Jin *et al.*, Science **360**, 893 (2018).

[27] H. Zeng, J. Dai, W. Yao, D. Xiao, and X. Cui, Nat. Nanotech. **7**, 490 (2012).

[28] K. F. Mak, K. He, J. Shan, and T. F. Heinz, Nat. Nanotech. **7**, 494 (2012).

[29] T. Cao *et al.*, Nat. Commun. **3**, 887, 887 (2012).

[30] A. M. Jones *et al.*, Nat. Nanotech. **8**, 634 (2013).

[31] X.-T. An, J. Xiao, M. W. Y. Tu, H. Yu, V. I. Fal'ko, and W. Yao, Phys. Rev. Lett. **118**, 096602 (2017).

[32] M. W.-Y. Tu and W. Yao, 2D Mater. **4**, 025109 (2017).

[33] G.-B. Liu, W.-Y. Shan, Y. Yao, W. Yao, and D. Xiao, Phys. Rev. B **88**, 085433 (2013).

[34] M.-H. Liu, J. Bundesmann, and K. Richter, Phys. Rev. B **85**, 085406 (2012).

[35] T. Ando, Phys. Rev. B **44**, 8017 (1991).

[36] W. Zhou, Y.-Y. Zhang, J. Chen, D. Li, J. Zhou, Z. Liu, M. F. Chisholm, S. T. Pantelides, and K. P. Loh, Sci. Adv. **4**, eaap9096 (2018).

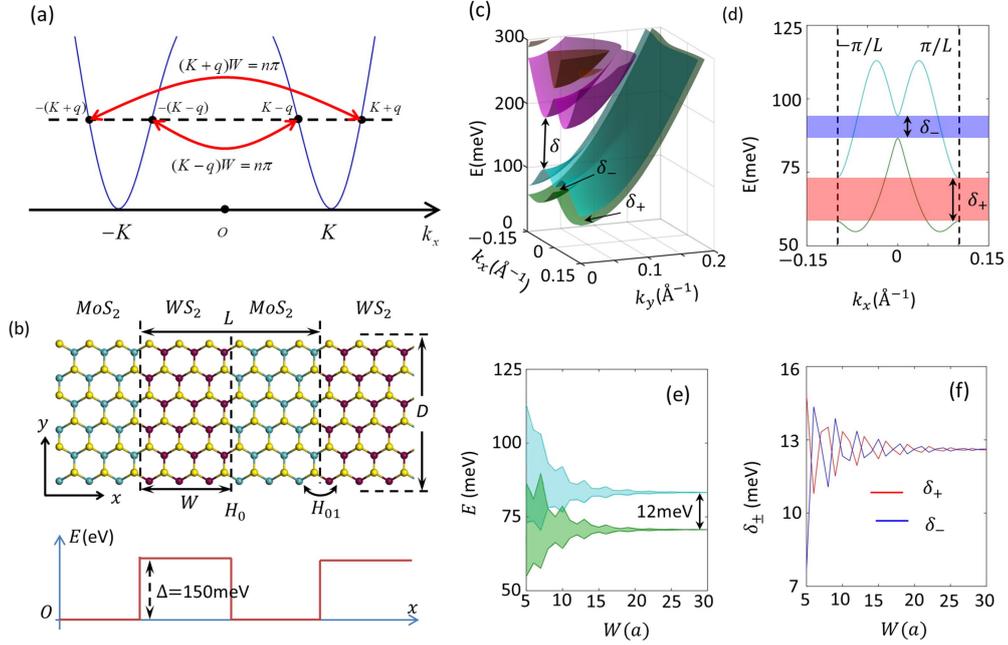

**Figure 1. (a)** The resonant tunneling conditions with the intervalley scattering. There are two kinds of intervalley resonant tunneling conditions, one is $(K+q)W = n\pi$, the other is $(K-q)W = n\pi$, where $q$ is the $x$ component of the wave vector measured from the valley center $\pm K$. For the superlattice, the intervalley resonant tunneling conditions can be give rise to valley-resolved miniband. **(b)** Schematic and band alignment for monolayer transition metal dichalcogenides lateral superlattice. **(c)** Miniband dispersion of $MoS_2/WS_2$ lateral superlattice with $L = 10a$, $W = 5a$, $a$ being the lattice constant of $MoS_2$. The gap $\delta$ is induced by the multiple intravalley reflection, gaps $\delta_+$ and $\delta_-$ by two kinds of intervalley reflection. **(d)** Zoom-in of the minibands in (c) for the two lowest bands, plotted at $k_y = 0$. Dashed vertical lines indicate minizone boundaries ($k_x = \pm\pi/L$). **(e)** The two lowest energies as functions of the barrier width $W$ for the fixed well width $L - W = 5a$. **(f)** $\delta_+$ and $\delta_-$ as functions of $W$, for the same superlattice periodicity.

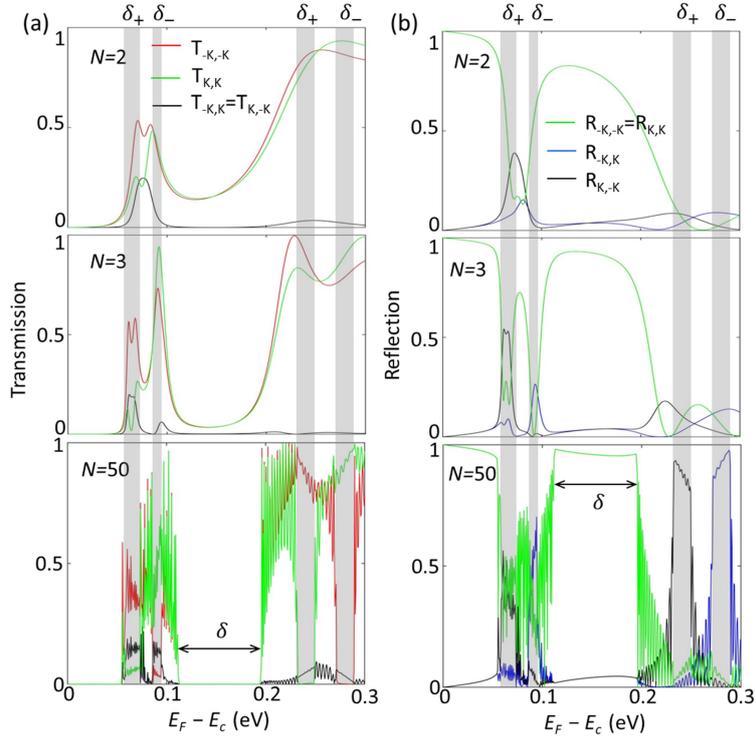

**Figure 2.** Valley-resolved transmission (a) and reflection (b) coefficients under normal incidence for periods of superlattice $N$=2, 3 and 50. $L = 10a$ and $W = 5a$ of such $MoS_2/WS_2$ heterojunction is used.

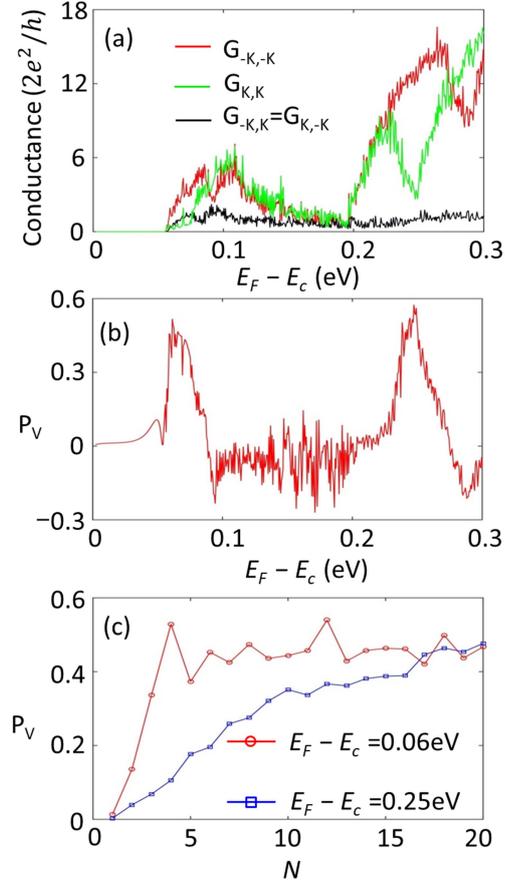

**Figure 3.** (a) Valley-resolved conductances as functions of the incident energy. (b) The valley polarization $P_V \equiv (G_{-K,-K} + G_{K,-K} - G_{K,K} - G_{-K,K})/G_{sum}$, where $G_{sum} \equiv G_{K,K} + G_{-K,-K} + G_{K,-K} + G_{-K,K}$ is the overall conductance. The calculations in (a) and (b) use a 20-period superlattice with $L = 10a$, $W = 5a$. (c) The valley polarization as a function of superlattice period $N$ for the various incident energies in the valley-resolved gaps. The width of the $MoS_2/WS_2$ heterojunction in *y* direction is chosen to be about 65nm.

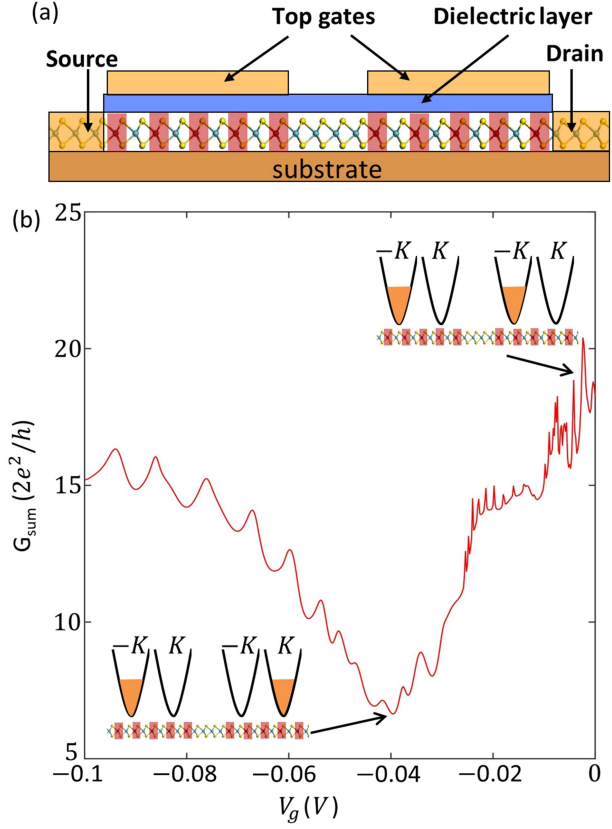

**Figure 4.** (a) Schematic diagram of the monolayer $MoS_2/WS_2$ giant valleyresistance device. The current through the device is blocked if two valley filters in series have opposite valley polarization. (b) The overall conductance for the device as a function of the gate voltage $V_g$ applied on the right valley filter. The chemical potential of the left valley filter $\mu_L = 0.247 eV$ is fixed in the gap $\delta_+$. $L = 10a$, $W = 5a$, and 20 periods of such superlattice is used. The width of the $MoS_2/WS_2$ valleyresistance device in $y$ direction is chosen to be about 65nm. The insets in (b) show the valley alignment.